\documentclass[sigconf]{acmart}
\pdfoutput=1
\usepackage{amsmath,amsfonts,amssymb}
\usepackage{eqnarray}
\usepackage{mathtools}
\usepackage{balance} 
\usepackage{mathrsfs}
\usepackage{multirow}
\usepackage{csquotes}
\usepackage{bbm}
\usepackage{enumitem} 

\usepackage[]{algorithm2e}
\usepackage{graphics}
\usepackage{subcaption}
\usepackage{units}
\usepackage{csquotes}
\usepackage{booktabs}
\usepackage{todonotes}

\copyrightyear{2017}
\acmYear{2017}
\setcopyright{rightsretained}
\acmConference{SIGIR 2017 Workshop on Neural Information Retrieval (Neu-IR'17)}{}{August 7--11, 2017, Shinjuku, Tokyo, Japan} \acmPrice{}
\makeatletter
\def\@fnsymbol#1{\ensuremath{\ifcase#1\or *\or \dagger\or \ddagger\or
   \mathsection\or \mathparagraph\or \|\or **\or \dagger\dagger
   \or \ddagger\ddagger \else\@ctrerr\fi}}
\makeatother

\title[DE-PACRR]
      {DE-PACRR: Exploring Layers Inside the PACRR Model}

\author{Andrew Yates}
\affiliation{%
  \institution{Max Planck Institute for Informatics}
}
\email{ayates@mpi-inf.mpg.de}

\author{Kai Hui}
\affiliation{%
  \institution{Max Planck Institute for Informatics}
  \institution{Saarbr\"ucken Graduate School of Computer Science}
}
\email{khui@mpi-inf.mpg.de}

\begin{abstract}
  Recent neural IR models have demonstrated deep learning's
  utility in ad-hoc information retrieval.
  However,
  deep models have a reputation for being black boxes, and the roles of a neural IR model's
  components may not be obvious at first glance. 
  In this work, we attempt to shed light on the inner workings
  of a recently proposed neural IR model, namely the PACRR model,
  by visualizing the output of intermediate layers and 
  by investigating the relationship between
  intermediate weights and the ultimate relevance score produced.
  We highlight several insights, hoping that such insights will be generally applicable.
\end{abstract}


\begin{document}
\maketitle

\section{Introduction} 
\label{sec.introduction}
The proposals of novel neural IR models~\cite{guo2016deep,mitra2017learning,hui2017position,DBLP:journals/corr/PangLGXC16}
have demonstrated deep learning's potential to advance ad-hoc information retrieval.
A better understanding of the functions and influences in practice of
different building blocks used in state-of-the-art neural
IR architectures may aid in further development of neural IR models.
In this work, we investigate the operation of the recently proposed PACRR model~\cite{hui2017position}
by visualizing and analyzing the model's weights after training.
In particular, we explore the roles of PACRR's pooling and combination layers by visualizing their
output and plotting relationships between their output and the final document relevance scores.
While doing so we highlight several insights which we deem to be important to the model's success,
with the hope that this will inspire the development of future models.
We remark that, while we hope these insights to be generally applicable, PACRR was developed for use
with data on the scale of traditional IR benchmark collections.
Our analyses were performed on such collections, and
thus our results are most applicable to this context.

 The rest of this paper is organized as follows.
 Section~\ref{sec.revisit} briefly describes the PACRR model.
 We introduce a running example and describe the datasets and hyperparameters used in Section~\ref{sec.visualization}
 before investigating the function of PACRR's layers in more detail.
 We conclude in Section~\ref{sec.conclusion}.

\section{Overview of PACRR}~\label{sec.revisit}
PACRR takes as input a similarity matrix between a query $q$ and a document $d$,
and outputs a scalar relevance score $\mathit{rel}(d, q)$
indicating the relevance between $q$ and $d$.
During training, one relevant and one non-relevant query-document pair
are encoded as similarity matrices.
The relevance scores for both documents are compared using a
max-margin loss as in Eq.~\ref{eq.maxmarginloss}.
\begin{equation}\label{eq.maxmarginloss}
 \mathcal{L}(q,d^+, d^-;\Theta)=\mathit{max}(0,1-\mathit{rel}(q,d^+)+\mathit{rel}(q,d^-))
\end{equation}
PACRR is composed of the following layers.
\begin{enumerate}
\item \textbf{Input}: $\mathit{sim}_{l_q\times l_d}$,
  where the query length $l_q$ and document length $l_d$ dimensions are fixed.
  That is, $\mathit{sim}_{i,j}$ contains the word2vec~\cite{word2vec} cosine similarity\footnote{We begin with the Google News word2vec embeddings and continue training them on our document corpus to avoid missing terms. We set the cosine similarity to 1 for terms that the Porter stemmer stems to the same strings.} between a query term at $i$ and a document term at $j$.
 \item \textbf{CNN kernels followed by max-pooling layers}: 
   multiple convolutional kernels with $l_f$ filters identify query-document term
   matches for different term window sizes, namely, $2, 3, \cdots, l_g$.
   The parameter $l_g$ determines the maximum kernel size.
   Afterwards, a max-pooling layer retains only the strongest filter signal for each kernel size,
 leading to $l_g$ matrices denoted as $$C^1_{l_q\times l_d \times 1}\cdots C^{l_g}_{l_q\times l_d \times 1}\;,$$
 which we call the \textit{filter-pooling} layer in this work.
 The matrix $C^1$ corresponds to the original similarity matrix, which already contains unigram scores.

 \item \textbf{A k-max-pooling} layer subsequently
 pools matching signals in $C^1,\cdots, C^{l_g}$,
 keeping only the top-$n_s$ strongest signals for each query term and kernel size pair.
 The output of this layer is
 $$P^1_{l_q\times n_s},\cdots, P^{l_g}_{l_q\times n_s}\;.$$
\item \textbf{Combining signals across query terms}.
  An LSTM layer processes the match signals for each query term, $P_{l_q\times (l_g n_s)}$,
  and outputs the document's final relevance score $\mathit{rel}(d, q)$.
\end{enumerate}

\section{Exploration}\label{sec.visualization}

\begin{figure*}[ht!]
  \includegraphics[width=0.8\textwidth]{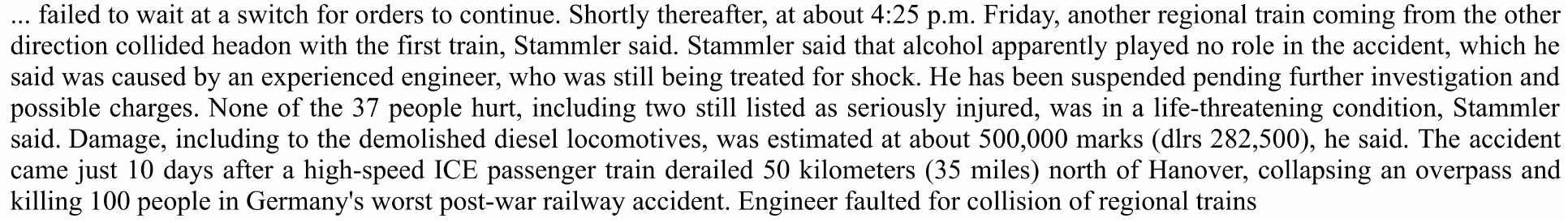}
  \caption{Snippet of relevant document APW19980613.0242 from Robust05.}
  \label{fig.doc}
\end{figure*}

\begin{figure*}[ht]
\centering
\begin{subfigure}[ht]{1\textwidth}
\centering
  \includegraphics[width=0.8\textwidth]{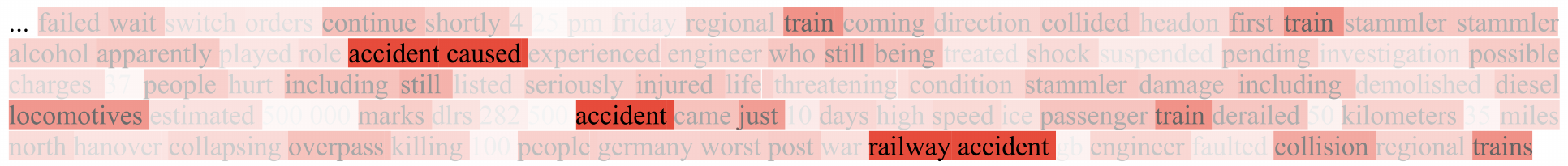}
  \caption{Text markup illustrating unigram term signals present after the filter-pooling layer.}
  \label{fig.unigram_all}
\end{subfigure}

\begin{subfigure}[t]{1\textwidth}
\centering
  \includegraphics[width=0.8\textwidth]{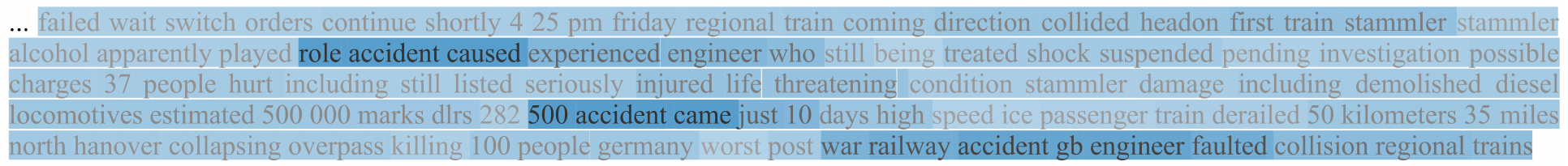}
  \caption{Text markup illustrating the 3x3 kernel signals present after the filter-pooling layer.}
  \label{fig.3gram_all}
\end{subfigure}

\begin{subfigure}[t]{1\textwidth}
\centering
  \includegraphics[width=0.8\textwidth]{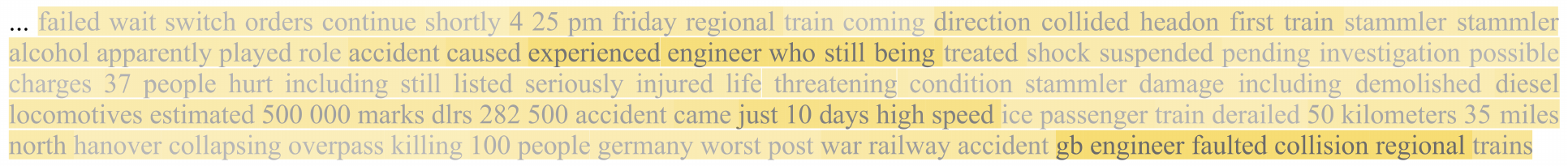}
  \caption{Text markup illustrating the 5x5 kernel signals present after the filter-pooling layer.}
  \label{fig.5gram_all}
\end{subfigure}
\caption{Text markup illustrating the output of the filter-pooling layer.}
\end{figure*}

As mentioned, 
the interactions between a query and a document are first encoded 
as the similarity matrix $\mathit{sim}_{l_q\times l_d}$.
Thereafter,
multiple kernels of different sizes are employed 
to extract salient matching signals locally, in line with practices in traditional ad-hoc retrieval models.
Next, \textit{filter-pooling} and \textit{k-max-pooling} layers
are used to retain the strongest signal(s)
for each kernel and each query term.
Ultimately,
these strongest signals are combined into a query-document relevance score $\mathit{rel}(d, q)$.
The PACRR model can be described as an extraction-distillation-combination sequence,
with CNN kernels extracting relevance matches, pooling layers distilling the matches into
a series of small vectors for each query term, and a final layer combining the query term signals
into an ultimate relevance score.
Given the large number and high dimensions of the signals extracted 
from the CNN kernels, only the distillation and combination steps are investigated in this work;
these are the steps in which the strongest relevance signals are identified and combined.
Following this extraction-distillation-combination framework, 
we attempt to better understand the functionality of 
different layers by visualizing their weights or by 
 correlating them with the model's ultimate output.

\textbf{Running example.}
  \begin{displayquote}
  Title: \textit{railway accidents}\\
  Description: \textit{what are the causes of railway accidents throughout the world?}\\
  Document: as displayed in Figure~\ref{fig.doc}
 \end{displayquote}
 
\textbf{PACRR Model.}
The model is trained over 200 Robust04 queries for 100 iterations and validated on the remaining 50 Robust04 queries.
The query-document pairs analyzed in this work are taken from Robust05.
We set $l_q=16$  and drop the lowest IDF terms after concatenating 
terms from the title and the description field in the 
queries from the \textsc{Trec} Robust Track\footnote{http://trec.nist.gov/data/robust.html}.
We use $l_g=5$ to enable $2\times2$, $3\times3$, $4\times4$ and $5\times5$ kernels.
The number of matching signals to keep for each query term is set to $n_s=10$.

\begin{figure*}
\centering
\begin{subfigure}[t]{1\textwidth}
\centering
  \includegraphics[width=0.8\textwidth]{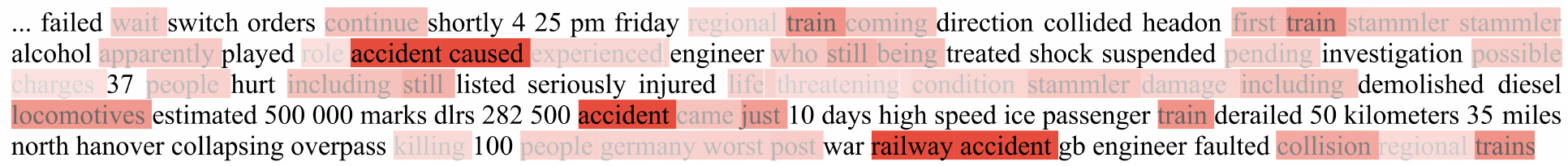}
  \caption{Text markup illustrating the remaining unigram term signals after the k-max-pooling layer.
  }
  \label{fig.unigram_topk}
\end{subfigure}

\begin{subfigure}[t]{1\textwidth}
\centering
  \includegraphics[width=0.8\textwidth]{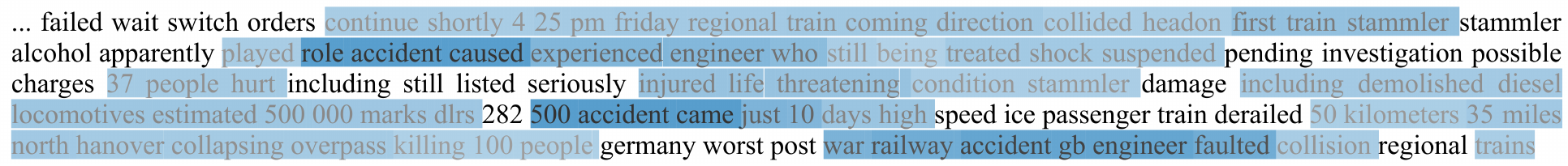}
  \caption{Text markup illustrating the remaining 3x3 kernel signals after the k-max-pooling layer.
  }
  \label{fig.3gram_topk}
\end{subfigure}

\begin{subfigure}[t]{1\textwidth}
\centering
  \includegraphics[width=0.8\textwidth]{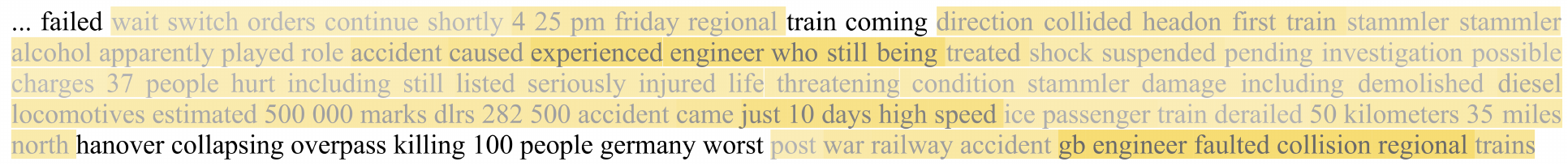}
  \caption{Text markup illustrating the remaining 5x5 kernel signals after the k-max-pooling layer.
  }
  \label{fig.5gram_topk}
 \end{subfigure}
 \caption{Text markup illustrating the output of the k-max-pooling layer.}
\end{figure*}

\textbf{Distillation.}
Two pooling layers are involved, namely, the \textit{filter-pooling} layer 
and the \textit{k-max-pooling} layer.

The use of a filter-pooling layer differs from the pooling strategies
employed in computer vision~\cite{simonyan2014very}, where
pooling layers serve to sub-sample different regions of an image.
PACRR's filter-pooling aims to retain only one salient signal
for each kernel among the different filters. The assumption is that all filters
measure different types of relevance matches, such as n-gram matches or term proximity matches,
thus only the strongest relevance signal needs to be kept.
This interpretation of the role of filters could be applied to any neural IR model
that performs relevance matching using a CNN.
To illustrate the signals that are distilled by this filter-pooling
layer, a snippet from the example document is displayed.
Figure~\ref{fig.unigram_all}, Figure~\ref{fig.3gram_all} and Figure~\ref{fig.5gram_all}
display the markup for kernels with sizes $1\times 1$, $3\times 3$ and $5\times 5$ respectively,
showing the strongest filter signal among all query terms. 
Kernels with other sizes, namely, $2\times 2$ and $4\times 4$,
are omitted given that similar patterns are observed.
The opacity (i.e., darkness of the text) represents the value of the output
of the filter-pooling layer, which is the strength of the signal.
The signal for each kernel size is normalized by the maximum value among all query-document pairs
to make relative differences in the visualization more clear.
In cases where there is overlap among different windows of text,
the strongest signal for each term is used to determine the term's opacity.

The text sequences with the darkest markup
represent the strongest signals, like the unigram signals in Figure~\ref{fig.unigram_all}
and the 3x3 signals in Figure~\ref{fig.3gram_all}.
Meanwhile, in Figure~\ref{fig.5gram_all}, 
a lighter color is observed even for the strongest signals, illustrating that 
the strength of the signals is generally smaller from a larger kernel.
The use of real valued cosine similarity in the input matrices
allows the model to match related terms beyond exact matches,
thereby expanding the query. For example, in Figure~\ref{fig.unigram_all}
the terms ``locomotives'' and ``collision'' have relatively high weights though neither term appears in the query.
We can also see that almost all terms have at least some weight
after the filter-pooling layer,
reducing the difference between
the salient text and the remaining text.
This is due to the way CNN kernels work when combined with real valued similarity.
Taking the dot product of all terms in a window generally produces non-zero values and acts as a smoothing effect.

\begin{figure*}[t]
    \centering
    \includegraphics[width=0.9\textwidth]{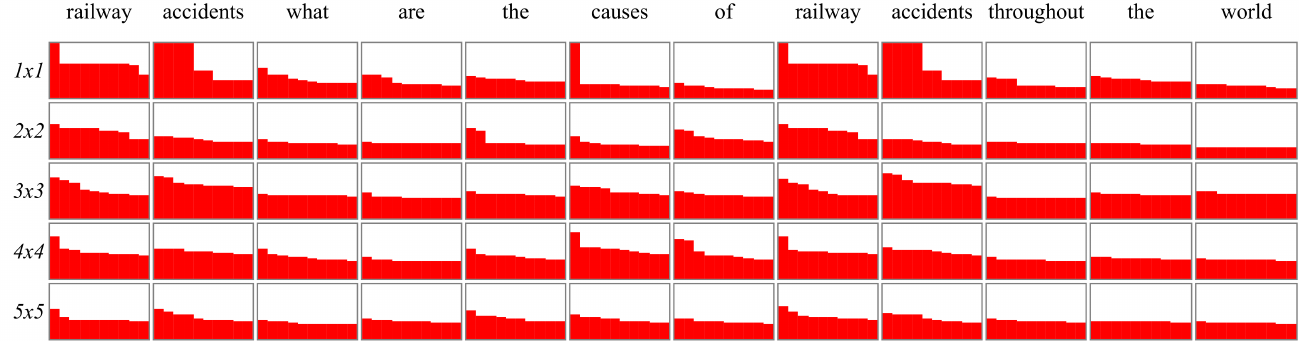}
    \caption{The complete output of the \textit{k-max-pooling} layer. Columns correspond to query terms.
    Rows correspond to kernel sizes (e.g., n-gram and term proximity matches). Each cell is composed of 10 bars indicating the strength of the
    top $n_s=10$ signals for the corresponding query term and kernel size.
    }
    \label{fig.qcrel}
  \end{figure*}

After the filter-pooling layer
a \textit{k-max-pooling} layer is employed to further retain the $n_s$-most salient signals
for each query term and kernel size pair,
allowing the later combination component to focus on only the strongest matches.
The use of k-max-pooling can be viewed as a trade-off between two extremes:
max pooling loses too much information about the number of matches for a given query term and
kernel size pair, whereas performing no pooling retains much information of minimal salience
and thus provides the combination layer with a noisier signal.
Any CNN-based relevance matching model can use any of these three pooling strategies;
which strategy is optimal likely depends on the training data available.
Markup figures after the k-max-pooling layer are displayed in
Figure~\ref{fig.unigram_topk}, Figure~\ref{fig.3gram_topk}
and Figure~\ref{fig.5gram_topk} for 
$1\times 1$, $3\times 3$ and $5\times 5$ kernels, respectively.
Compared with the corresponding figures for the filter-pooling layer,
the k-max-pooling layer has removed most of the ``background'' term signals.
Analogous to a user who finds key terms in a web page,
the layer's output focuses on the few most relevant text sequences rather than considering everything.
To better understand its functionality, we further summarize 
the output of this layer in Figure~\ref{fig.qcrel}, 
which visualizes a complete example output of the k-max-pooling layer. 
Each column corresponds to a query term and each row corresponds to one kernel size.
Each cell is composed of ten bars displaying the strength for the top $n_s=10$
signals for each query term under a particular kernel size.
As in the earlier figures, the strength of the signals are normalized to aid in visualization.
It can be seen that the distribution of the top-$n_s$ signal strengths
vary widely among different query terms and among kernel sizes.
Smaller kernels are more likely to have stronger matches, however, and in the next section
we demonstrate that the combination layer learns to account for this.

We argue that the salient signals under a kernel with size $l$x$l$ are a mixture of 
$l$-gram matching and query proximity in a small text window with $l$ terms.
The latter kind of signals account for more of the signals with larger $l$, such as 3x3 kernels.
For example, the text sequence ``role accident caused''
from Figure~\ref{fig.3gram_topk} is highlighted because
it contains\footnote{``Causes'' and ``caused'' are equivalent after stemming.} the query term ``causes,'' not because it is a query trigram.
Interestingly, this match was identified by a 3x3 kernel, yet there is no 3-term query window
containing both ``accident'' and ``causes.'' In this match the terms ``role'' and ``accident'' have high
weight because they have relatively high word2vec similarity with ``causes,'' not because
they are matching other query terms. That is, the two query terms ``accident'' and ``cause'' are too far
away from each other to both be considered by the same 3x3 kernel, and thus the high weight given to
``role accident caused'' comes from each term's relatively high similarity to the single query term ``causes.''
We note that this behavior stems from PACRR's use of word2vec embeddings to calculate term similarity,
thus it should apply to any model that uses term embeddings rather than exact matching.

\begin{figure*}
  \label{fig.topk}
  \centering
  \begin{subfigure}[t]{0.48\textwidth}
    \centering
    \includegraphics[width=\textwidth]{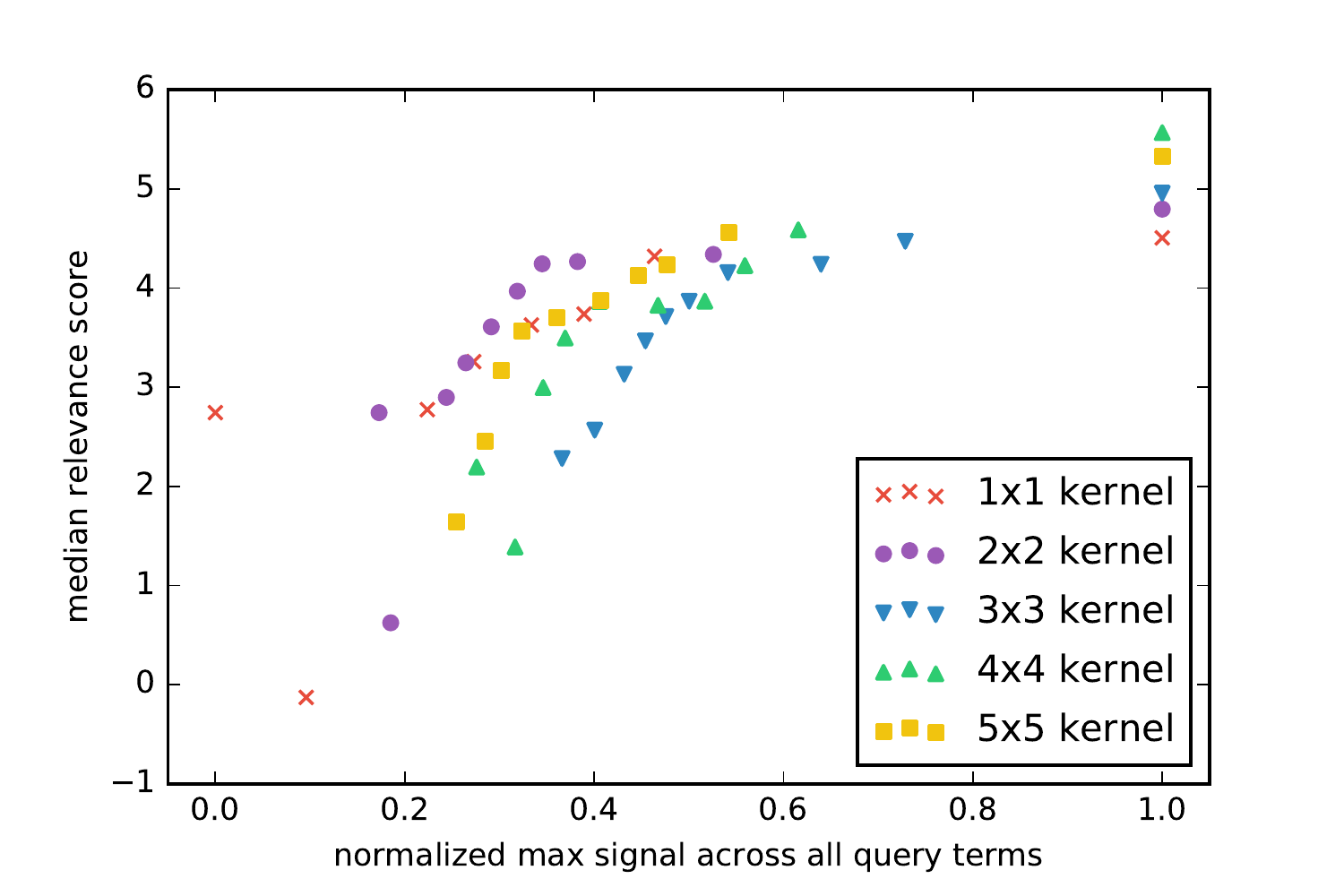}
    \caption[]{{\small Strongest signals in the top-k (i.e., top-k position one)}}
    \label{fig.topk1}
  \end{subfigure}
  \begin{subfigure}[t]{0.48\textwidth}
    \centering
    \includegraphics[width=\textwidth]{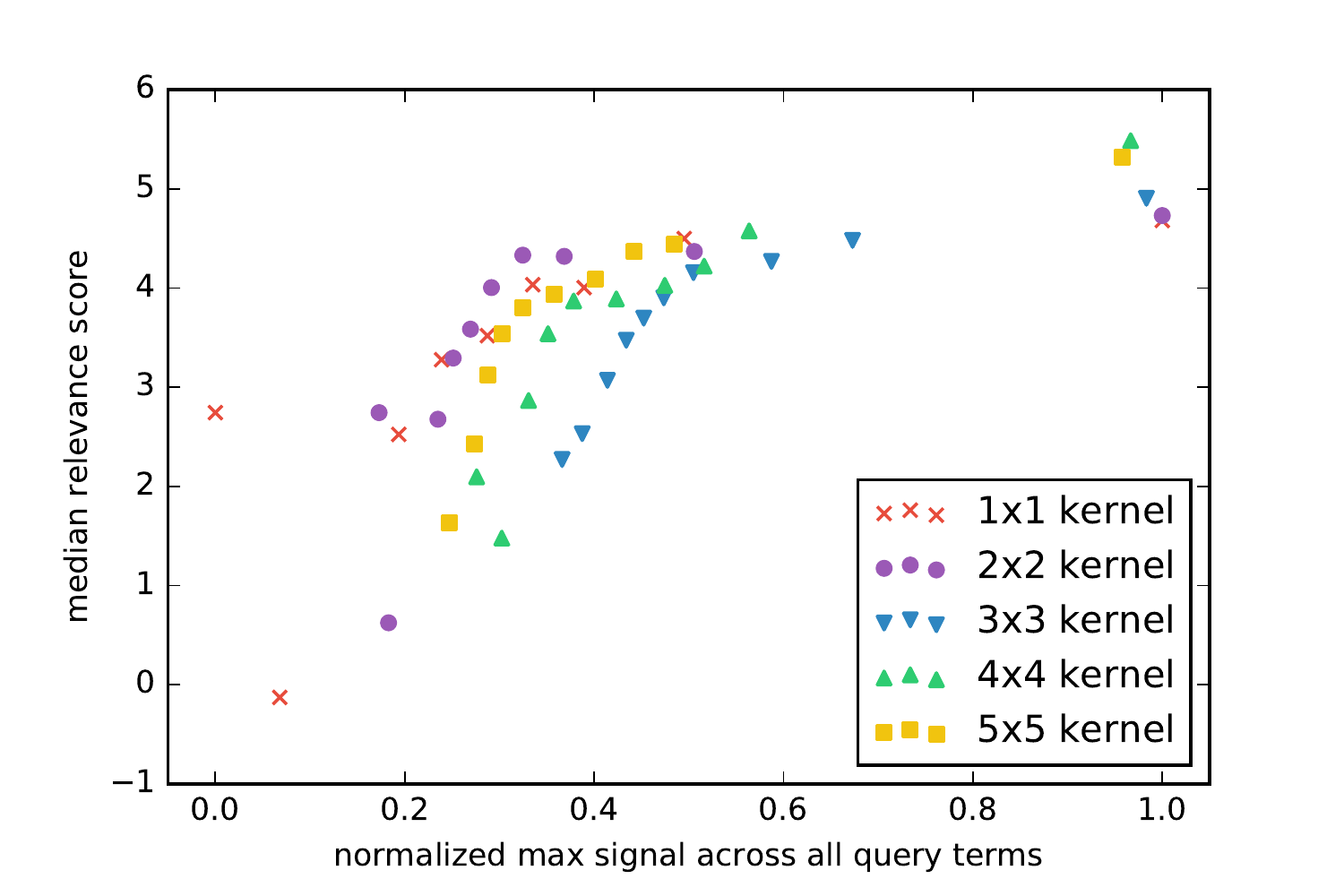}
    \caption[]{{\small Second strongest signals in the top-k (i.e., top-k position two)}}
    \label{fig.topk2}
  \end{subfigure}
  \begin{subfigure}[t]{0.48\textwidth}
    \centering
    \includegraphics[width=\textwidth]{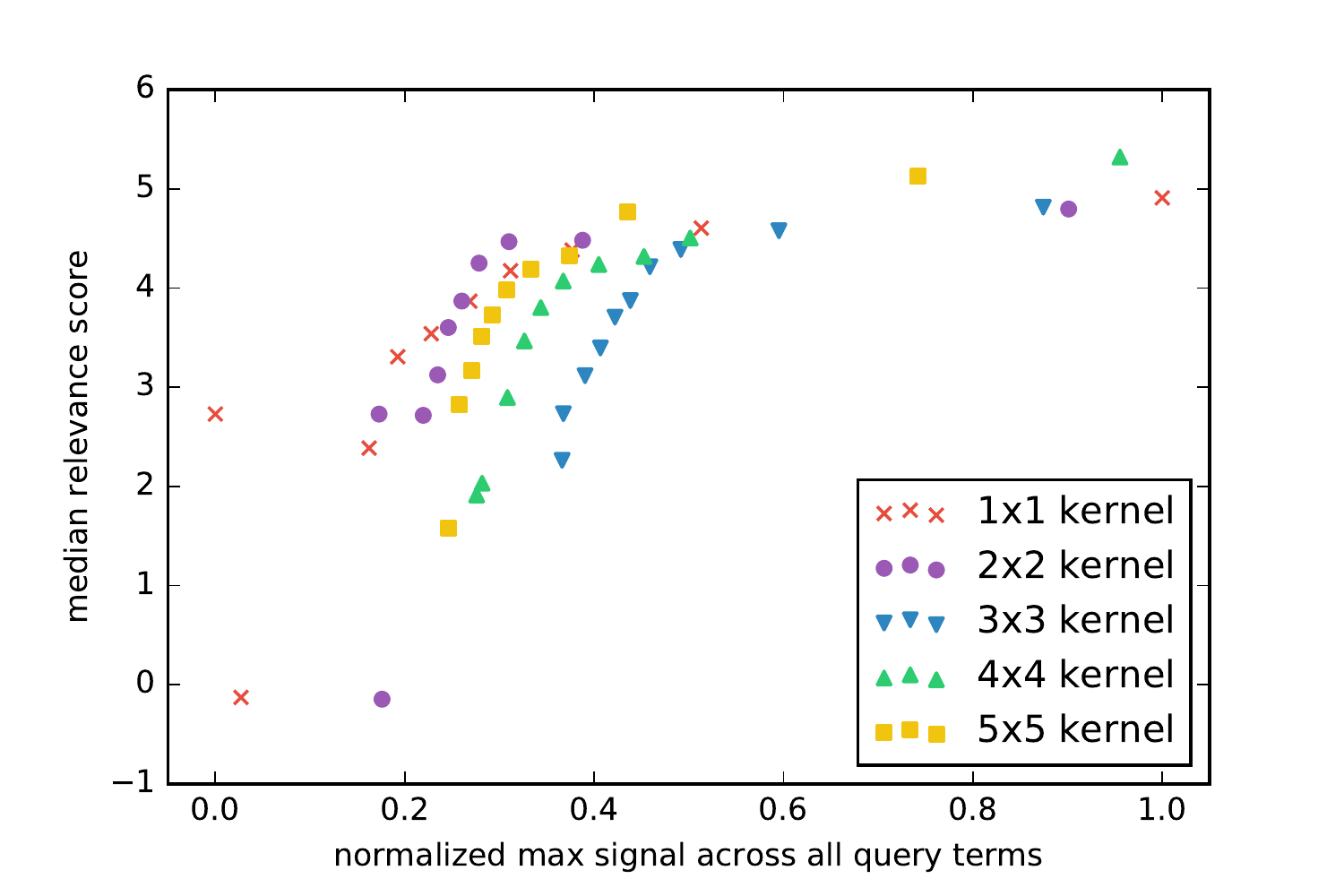}
    \caption[]{{\small Fifth strongest signals in the top-k (i.e., top-k position five)}}
    \label{fig.topk5}
  \end{subfigure}
  \begin{subfigure}[t]{0.48\textwidth}
    \centering
    \includegraphics[width=\textwidth]{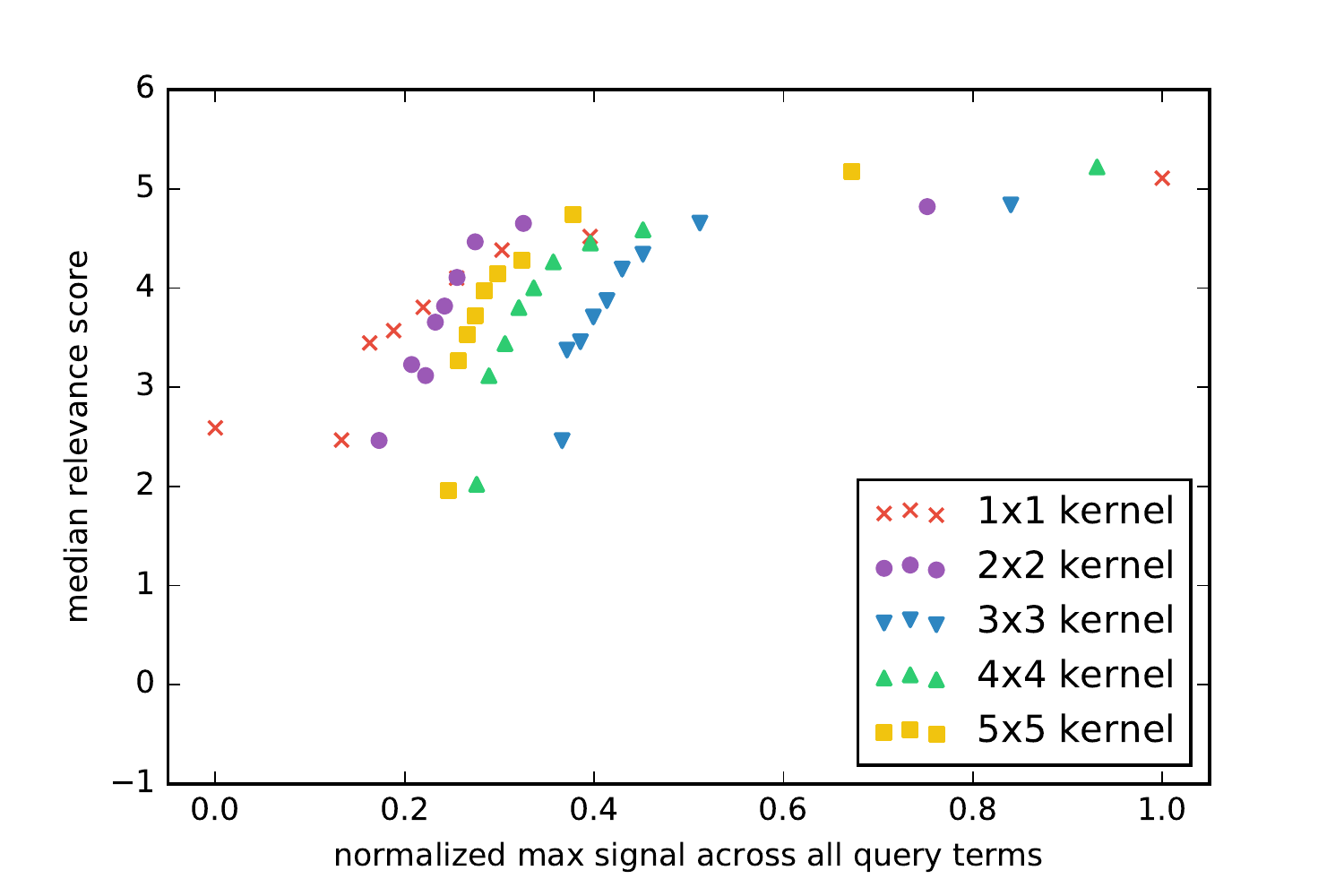}
    \caption[]{{\small Tenth strongest signals in the top-k (i.e., top-k position ten)}}
    \label{fig.topk10}
  \end{subfigure}
  \caption[]{The relationship between documents' signal strengths and documents' relevance scores
  for different kernel sizes and positions in the top-k. The difference in scores between kernel sizes increases as the top k position increases.}
  \label{fig.topk}
\end{figure*}


\textbf{Combination.}
After extracting the k-most salient signals for each kernel along different query terms,
the model combines them into a document relevance score $\mathit{rel}(d, q)$.
Given the large number of weights involved in the combination layer,
we investigate the relationships between different signal types and the relevance score.
The combination procedure can be viewed as a function mapping the salient signals
from the previous step to a real value.
As displayed in Figure~\ref{fig.qcrel},
the combination step's input consists of the top $n_s=10$ signals for different query terms and kernel sizes;
this Figure illustrates the combination layer's entire input.

In this section, we consider the following questions:
\begin{itemize}
 \item[-] How are signals from different kernels combined?
 \item[-] How are signals from different top-k positions combined?
\end{itemize}
To do so, we consider the signals for each position in the top-k one at a time
(e.g., we consider only the second strongest signals).
For each position in the top-k and each kernel size, we divide all signals from the query terms
into ten bins of equal sizes.
For each bin we report the median of the ultimate relevance score produced by the combination layer.
This relationship between signals and relevance scores is illustrated in
Figure~\ref{fig.topk}, where
the x-axis corresponds to the strength of the signals for different bins, 
and the y-axis is the median of the final relevance score.

Figure~\ref{fig.topk} illustrates the fact that different kernel sizes are weighted differently
by the combination layer. For example, 
in the upper right corner of Figure~\ref{fig.topk10},
the strongest unigram match with a strength of 1.0 leads to lower
relevance scores than the strongest 5x5 match with a strength of only $0.7$.
One explanation is that 
the loss function in Eq.~\ref{eq.maxmarginloss}
compares a relevant and a non-relevant document, which both can include similar amounts of unigram matches,
making the contributions of the unigram signals less important.
Intuitively, even after a document includes all separate query terms,
its relevance score can still benefit from considering other factors, such as
the relevance signals produced by 2x2 or 3x3 kernels.
Strong 5x5 signals are more rare, 
thus the combination layer tends to reward a document more 
when such rare signals are observed.
Additionally, Figure~\ref{fig.topk} contains clear outliers in the leftmost region:
there are some documents that have only weak unigram matches,
but still receive a relevance score of approximately 2.8. This illustrates the weight that the model
gives to inexact term proximity matches from larger kernels which also include neighboring terms.

Regarding the second question, Figure~\ref{fig.topk} indicates
that all signals in the top-k are considered
when combining results. This illustrates the utility of performing k-max pooling rather than max pooling,
as is commonly done in computer vision. 
For example, in Figure~\ref{fig.topk5}, the fifth strongest signals for 
the $5\times 5$ kernel are always less than 0.8, 
but the corresponding relevance score is still as large as the highest relevance score in that figure.
Put differently, 
though the absolute values of the matching scores decrease when considering lower ranked signals,
e.g., a 2x2 kernel's maximum signal is approximately 1.0 in the 2nd position and 0.7 in the 10th position,
such later positions still contribute strongly to the ultimate relevance score.
This consideration of all of the top-k signals is analogous to the computations employed in many
traditional IR methods, such as TF-IDF, where all occurrences of the query terms are aggregated.

\section{Conclusion}\label{sec.conclusion}
In this work we explored the pooling and the combination layers from the recently proposed
PACRR model, aiming at generally applicable insights.
We notice that the real valued similarity
from the usage of word2vec
expands the query,
allowing the model to assign weights to windows of text with little or even no exact query overlap.
Together with the usage of kernels with different sizes,
the real valued similarity further enables
proximity matching, which becomes more common as the kernel size (i.e., window length) increases.
Subsequently,
different pooling layers retain the strongest signals from these kernels,
making the model focus on the most salient matches.
At the time of combination,
such signals from different kernels with different strengths 
are comprehensively considered by the model, 
highlighting the necessity to retain more than 
the top-1 signal in the pooling layer.
Moreover, we remark that
the combination layer actually emphasizes
the signals from larger kernel sizes more strongly,
given their rarity relative to the unigram signals.
This demonstrates the strength of a neural IR model to go beyond unigram matches.

\balance
\bibliographystyle{ACM-Reference-Format}
\bibliography{kai}  

\end{document}